\documentclass[10pt,a4paper]{llncs}
\usepackage[a4paper, top=5.2cm, bottom=5.2cm, left=4.4cm, right=4.4cm]{geometry}
\usepackage{mathptmx} % Times New Roman
\usepackage{microtype}
\usepackage[T1]{fontenc}
\usepackage[utf8]{inputenc}
\usepackage{csquotes}
\usepackage{float}
\usepackage{graphicx}
\usepackage{amsmath}
\usepackage{amssymb}
\usepackage{listings}
\usepackage{braket}
\usepackage{xcolor}
\usepackage[style=apa]{biblatex}
\usepackage{hyperref}

\begin{document}
% Line break (\\) removed to avoid hyperref warning in PDF string
\title{GroverFigureOfMerit: An Agnostic Figure of Merit for Quantum Backend Characterization in the NISQ Era}
\author{Tiago Restucha\inst{1}\and Marcos Guillermo Lammers\inst{1}\and Alejandro  Fernández\inst{1}}
\authorrunning{T. Restucha and M.G. Lammers and A. Fernández}
\institute{Universidad Nacional de La Plata, Facultad de Informática, Laboratorio de Investigación y Formación en Informática Avanzada (LIFIA), Argentina}
\maketitle
\begin{abstract}
The Noisy Intermediate-Scale Quantum (NISQ) era poses a concrete challenge for developers: quantum hardware providers expose their capabilities through heterogeneous interfaces with proprietary metrics that vary widely across providers, hindering informed backend selection. The commonly available static characterization metrics—coherence times $T_1$/$T_2$, gate error rates, among others—exhibit specific limitations: they fail to capture dynamic variability across successive executions, overlook the impact of the transpilation process, and lack architectural comparability across physically distinct technologies.
In this work, we propose a Figure of Merit (FoM) based on Grover's algorithm, designed to act as an algorithmic stress test that holistically evaluates the performance of quantum backends. The proposed metric combines the probability of success on target states with penalties for non-uniform amplification and probability leakage to non-marked states, yielding a unified and comparable score across different hardware architectures. We implement this FoM on top of the Qonscious framework, a conditional execution platform that readily provides the polymorphic adapters needed to execute the same metric in an agnostic manner on IBM and IonQ backends, as well as simulators.
The main contributions are: (1) the proposal and validation of the GroverFigureOfMerit, a Grover-based metric that incorporates uniformity and leakage penalties (adapted from previous approaches such as GRADE), with emphasis on capturing the combined effect of noise, transpilation, and topological constraints; (2) a systematic analysis of heterogeneity across nine quantum providers, which motivates the need for agnostic metrics; and (3) the experimental demonstration of the proposed FoM through executions on ideal simulators and noise models derived from real processors, demonstrating its sensitivity to noise, topology, and transpilation overhead. The results confirm that the proposed metric enables distinguishing backend performance under a unified score, capturing intrinsic algorithmic limits. Validation on physical QPUs is identified as a natural next step.
\keywords{NISQ computing \and quantum benchmarking \and Grover's algorithm \and GRADE \and backend characterization \and Qonscious framework}
\end{abstract}
\section{Introduction}
The proliferation of diverse quantum computing architectures demands a robust and standardized methodology for evaluating the performance of \textit{backends}. Quantum computers are complex systems whose emergent behavior cannot be captured or predicted reliably based solely on low-level component metrics \parencite{proctor2024}. Raw device parameters prove insufficient due to their context-free nature, temporal instability, and lack of architectural comparability across providers such as IBM, IonQ, and Rigetti.
Consequently, this work proposes a Figure of Merit (FoM) for quantum backend characterization. Specifically, we designed the GroverFigureOfMerit, a metric that uses Grover's algorithm as an algorithmic \textit{stress test} to capture the combined effect of noise, transpilation, and topological constraints in a unified score. To address heterogeneity in execution, we implemented our FoM on top of the Qonscious framework \parencite{lammers2025}, incorporating the scoring function from recent approaches such as GRADE \parencite{manor2025} into its extensible architecture.
\subsection{Static Metrics vs. Algorithmic Performance}
Provider-specific characterization metrics are valuable for understanding hardware at the component level. However, their limited predictive power has driven an increasing focus toward benchmarks that summarize the system's holistic performance in a direct and concise manner \parencite{lubinski2023}. From our perspective, this evolution toward holistic benchmarks responds to the fact that low-level metrics, when analyzed in isolation, face the following particularities:
\textbf{Limited predictive power:} A backend with low gate error rates may fail to amplify target states in algorithms such as Grover's search. This occurs because low-level metrics do not account for the impact of the transpilation process, which generates a physical depth that is often substantially greater than the original high-level circuit depth. For example, multi-controlled gates (such as Toffoli gates) must be decomposed into multiple native one- and two-qubit gates to enable execution, introducing excessive circuit depth that compromises overall performance \parencite{lubinski2023}.
\textbf{Lack of contextual relevance:} Static metrics fail to encapsulate the behavior of particular algorithms. A set of gates with high individual fidelity may not translate into a successful execution of a specific algorithm if the interaction between gates, idle times (\textit{idle}) between operations, and topology affect the final result in a non-trivial way.
\textbf{Temporal variability:} Hardware properties change constantly due to environmental factors and the accumulation of random perturbations from the physical environment. As evidenced in the literature, this ``unpredictable frequency distribution and temporal variability interferes with qubit manipulation, making frequent recalibration necessary'' \parencite{jiang2025}. These continuous recalibrations cause static characterization snapshots to quickly lose validity, offering a blurry view of the backend's actual state at the time of execution.
These characteristics motivate a complementary approach: \textit{circuit-based FoMs} that measure end-to-end performance by executing quantum circuits and analyzing the results. Unlike static metrics, FoMs are:
\begin{itemize}
    \item \textbf{Dynamic:} They capture real execution behavior including transpilation effects, noise accumulation, and algorithm-specific performance.
    \item \textbf{Backend-agnostic:} The same FoM circuit can be executed on any quantum backend regardless of the provider's specific APIs.
    \item \textbf{Interpretable:} FoMs produce quantitative scores that directly relate to algorithmic success criteria.
\end{itemize}
\subsection{Contributions}
The main contributions are the heterogeneity analysis across nine providers, the extension of the Qonscious framework with backend-agnostic circuit-based FoMs, and the experimental validation of the proposed metric on ideal simulators and realistic noise models.
\subsection{Related Work}
Benchmarking in the NISQ era has evolved from low-level static metrics toward holistic algorithmic evaluations that capture real end-to-end performance \parencite{lubinski2023, proctor2024}. Tools such as Benchpress \parencite{Nation2025} offer varied suites of representative circuits, while GRADE introduces a Grover-based approach \parencite{manor2025}. Frameworks such as QUARK and BenchQC emphasize extensibility and portability; in our case, we emphasize the integration of metrics directly at runtime for resource-aware execution \parencite{lammers2025}.
\section{Heterogeneity Analysis Across Multiple Backends}
\begin{figure}[H]
    \centering
    \includegraphics[width=1\linewidth]{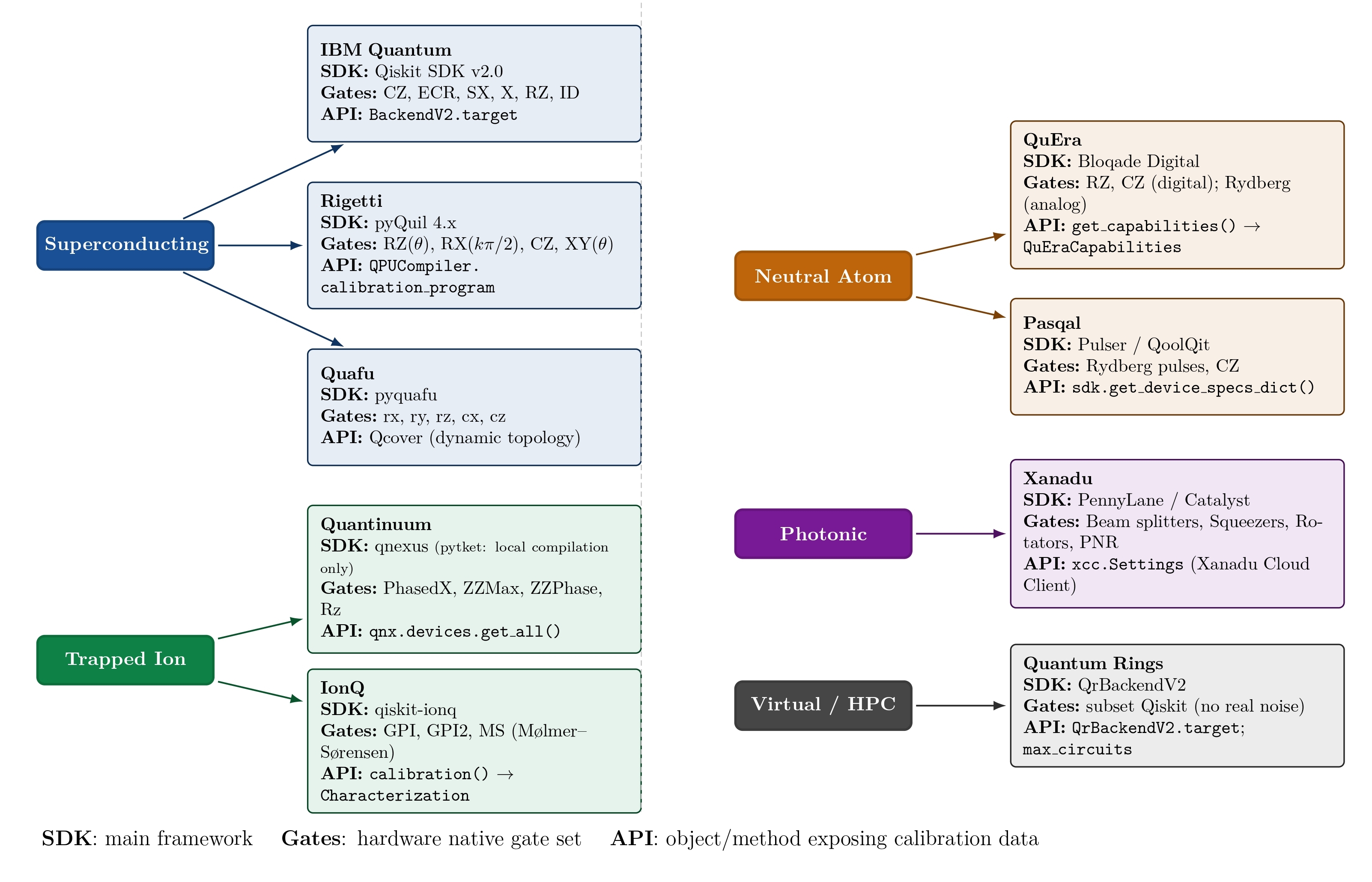}
    \caption{Classification of quantum computing providers by physical architecture.}
    \label{fig:proveedores}
\end{figure}

To gauge the complexity faced by a developer in the NISQ era, Figure~\ref{fig:proveedores} presents a taxonomy of providers grouped by their physical architecture. Each provider imposes its own SDK, its own set of native gates (\textit{gate set}), and its own mechanisms for exposing calibration data.
This difference is logical since ``different physical realizations of qubits give rise to distinct circuit primitives and computational trade-offs'' but the reality is that ``this circuit-based paradigm is largely hardware-agnostic and maps naturally onto most major quantum processor technologies, making it the dominant abstraction for quantum computing in the NISQ era''\parencite{gonski2026}. Therefore, at the algorithmic layer it should be practically invisible. The analysis of calibration interfaces reveals three barriers that prevent predicting circuit success based solely on static metrics:
\begin{enumerate}
    \item \textbf{API and Native Gate Fragmentation:} Even within the same physical architecture, differences exist. In superconducting systems, \textcite{api_ibm_2026} and \textcite{api_rigetti_2026} use distinct native gates (ECR/SX versus CZ/CPHASE) and expose topology through different types of objects (JSON, data dictionaries, etc.). This same interface heterogeneity is observed throughout the quantum ecosystem, from trapped-ion platforms to photonic systems \parencite[see for example,][]{api_ionq_2026, api_xanadu_2026}. From our perspective, designing portable compilers becomes a tedious problem.

    \item \textbf{Semantics of Metrics:} Low-level calibration data is generally not comparable. IBM reports gate errors and durations for each specific physical link. In contrast, architectures from \textcite{api_quantinuum_2026} or neutral atoms from \textcite{api_quera_2026} typically report global metrics or averages, given that their qubits are identical or dynamically reconfigurable.

    \item \textbf{Opacity and Volatility:} Quantum noise is not static, but the availability of dynamic data is uneven. Some providers expose rich characterization histories, while others offer limited or heavily encapsulated access, complicating any automated hardware selection workflow.
\end{enumerate}
From the developer's perspective, the relevant variables are the effective constraints of the computational graph (width, supported depth, and real connectivity cost). Consequently, this work adopts a top-down approach in favor of standardized FoMs based on algorithmic execution. This approximation allows:
\begin{itemize}
    \item To abstract proprietary API calls through polymorphic adapters, enabling the same benchmark to be executed on any backend.
    \item To evaluate real hardware performance through dynamic FoMs that capture transpilation, noise, and topology effects.
    \item To characterize processor suitability for specific computational tasks through algorithm-centric metrics, thus translating physical disparity into a comparable and agnostic score (\textit{score}).
\end{itemize}
\section{The Qonscious Framework as an Implementation Platform}
\begin{figure}[htbp]
    \centering
    \includegraphics[width=0.75\linewidth]{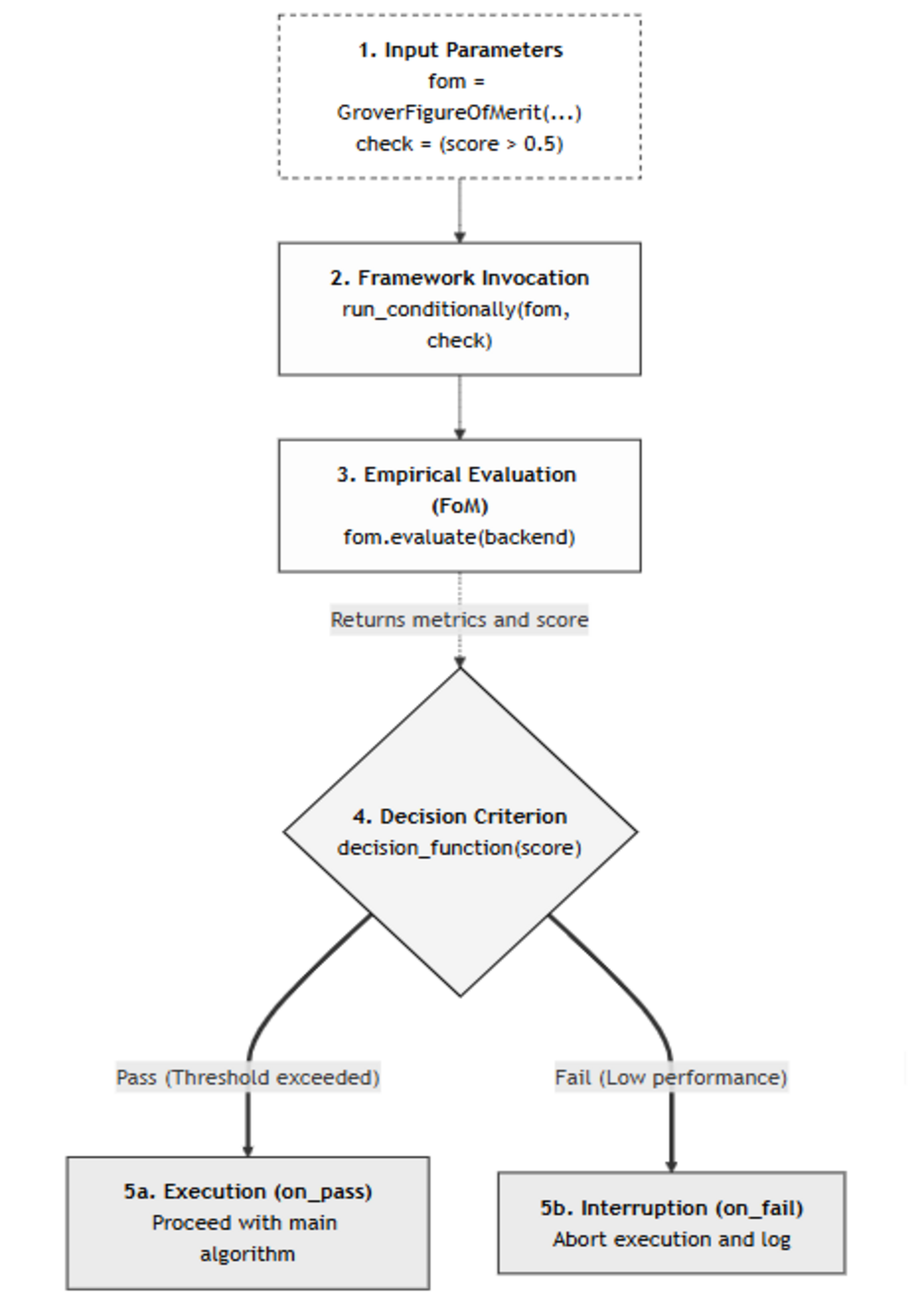}
    \caption{Qonscious conditional execution flow. The FoM result (score) can feed the decision function to optimize quantum resource usage.}
    \label{fig:flujo_qonscious}
\end{figure}

The proposed FoM has been implemented on the Qonscious framework \parencite{lammers2025}, an open-source platform originally designed for resource-aware conditional execution in NISQ environments. Qonscious provides two elements that facilitate the implementation of backend-agnostic FoMs:
\begin{itemize}
\item \textbf{Polymorphic adapters (BackendAdapters):} They abstract differences between SDKs and APIs from different providers. The same logical circuit can be transpiled and executed on any supported backend without modifying the FoM code.
\item \textbf{Conditional execution flow:} It allows parametrizing checks whose results feed a decision function that can trigger, pause, or cancel the main circuit execution. This flow, illustrated in Figure~\ref{fig:flujo_qonscious}, is what we use to execute our metric as a characterization mechanism.
\end{itemize}
In this work we focus on the characterization phase, using Qonscious's flow to execute the FoM on multiple backends and obtain comparable scores. The integration of these scores as conditions for runtime decisions is enabled for future work.

\section{GroverFigureOfMerit: Design and Definition}
\label{sec:grover-fom}
To demonstrate the practical utility of the proposed approach, we implemented a FoM based on Grover's algorithm, adapting the scoring function from GRADE \parencite{manor2025} to the extensible architecture of Qonscious.

\subsection{Motivation and Design}
Grover's algorithm \parencite{grover1996} quadratically amplifies the probability of measuring target states in an unstructured search space. In an ideal noiseless backend, Grover achieves a success probability close to 1 for the optimal number of iterations $R = \lfloor \frac{\pi}{4}\sqrt{N/M} \rfloor$, where $N$ is the size of the search space and $M$ is the number of target states.

On NISQ devices, noise degrades this amplification. The depth of the Grover circuit scales with $R$, and each iteration includes multi-controlled operations (\textit{oracle}) and multiple 1- and 2-qubit gates (\textit{diffusion}). This makes Grover an ideal algorithmic \textit{stress test} for evaluating hardware under real-world algorithmic execution conditions.

\subsection{Circuit Construction}
The Grover circuit is constructed in four stages:

1. Initialization: Apply Hadamard gates to the $n$ qubits to create a uniform superposition:
\begin{equation}
    |\psi\rangle = \frac{1}{\sqrt{N}} \sum_{x=0}^{N-1} |x\rangle
\end{equation}

2. Oracle $O_\omega$: Marks the solution by applying a phase inversion:
\begin{equation}
    O_\omega |x\rangle =
\begin{cases}
- |x\rangle, & \text{if } x = \omega, \\
\ \ |x\rangle, & \text{otherwise.}
\end{cases}
\end{equation}

3. Diffusion Operator $D$: Inverts the amplitudes with respect to the mean:
\begin{equation}
    D = 2|\psi\rangle\langle\psi| - I
\end{equation}

4. Iteration: Repeat $R$ times, where the optimal number is approximately:
\begin{equation}
    R^* \approx \left\lfloor \frac{\pi}{4} \sqrt{\frac{N}{M}} \right\rfloor
\end{equation}

with $M$ being the number of marked states.

\subsection{Score Calculation and Parameterization}
After running the circuit with $S = 2000$ repetitions (\textit{shots}), we obtain a count distribution $\{c_x\}$ for each bitstring $x \in \{0, 1\}^n$. We estimate the relative measurement frequencies as $f(x) = c_x / S$, which approximate the probability distribution of the final quantum state. For score calculation purposes, we use $P(x) \approx f(x)$ as an estimator for the probability of each bitstring.

\begin{itemize}
    \item Cumulative relative frequency on targets, where $T$ is the set of target states:
    \begin{equation}
        P_T = \sum_{x \in T} f(x)
    \end{equation}
    
    \item Probability on non-targets:
    \begin{equation}
        P_N = 1 - P_T = \sum_{x \notin T} P(x)
    \end{equation}
    
    \item Standard deviation among targets:
    \begin{equation}
        \sigma_T = \sqrt{\frac{1}{M}\sum_{x \in T}(P(x) - \bar{P}_T)^2}, \quad \bar{P}_T = \frac{P_T}{M}
    \end{equation}
    This metric penalizes situations where one target is amplified significantly more than others (indicating an imbalance caused by coherent errors).
    
    \item Final score:
    \begin{equation}
        \text{Score} = \max(0, P_T - \lambda \cdot \sigma_T - \mu \cdot P_N)
    \end{equation}
\end{itemize}

For circuit construction and subsequent evaluation, the implementation defines the metric through a set of configurable parameters. First, the number of states to mark is set via \texttt{num\_targets} ($M$). The analytical behavior is tuned with \texttt{lambda\_factor} ($\lambda$) and \texttt{mu\_factor} ($\mu$), which introduce penalties for probability imbalance among the targets themselves and for dispersion toward unwanted states, respectively. Additionally, the system accepts optional configurations such as \texttt{num\_qubits} ($n$), which fixes the width of the quantum register and is automatically inferred from $M$ if omitted, and \texttt{targets\_int}, which allows assigning an explicit list of integers representing the target states instead of delegating their selection to a random process. If $\mu \cdot P_N \geq P_T$, then Score = 0. This reflects situations where transpiled circuits increase in depth due to the decomposition of multi-controlled gates. Typical values for the balancing parameters are $\lambda = 1$ and $\mu = 1$ to equally penalize non-uniformity and probability dispersion, although they can be adjusted according to the application context or the characteristics of the evaluated backend.

\section{Experiments and Results}
To empirically validate the sensitivity of the proposed indicator to different hardware constraints, a comparative characterization experiment was designed using the native adapters of the Qonscious \textit{framework}. The goal is not to establish a definitive \textit{benchmark} of commercial state-of-the-art systems, but rather to demonstrate that the metric is sensitive to noise, topology, and transpilation overhead in a differentiated manner across backends. For this purpose, the employed noise models ---the IBM \textit{Fake Providers} family and the IonQ \textit{Aria 1} model--- constitute a sufficient and controlled validation baseline: they are models derived from real hardware characterizations that preserve the topological properties and error profiles of their corresponding physical processors, enabling experimental reproducibility without relying on cloud QPU availability. Validation on real physical processors, which would allow capturing additional dynamic phenomena such as \textit{crosstalk}\footnote{\textit{Crosstalk} refers to unwanted interference between adjacent qubits during simultaneous gate execution, causing correlated errors that are not captured by independent noise models.} and thermal fluctuations, is identified as a natural next step for confirmation in future work.

\begin{figure}[H]
    \centering
    \includegraphics[width=0.8\textwidth]{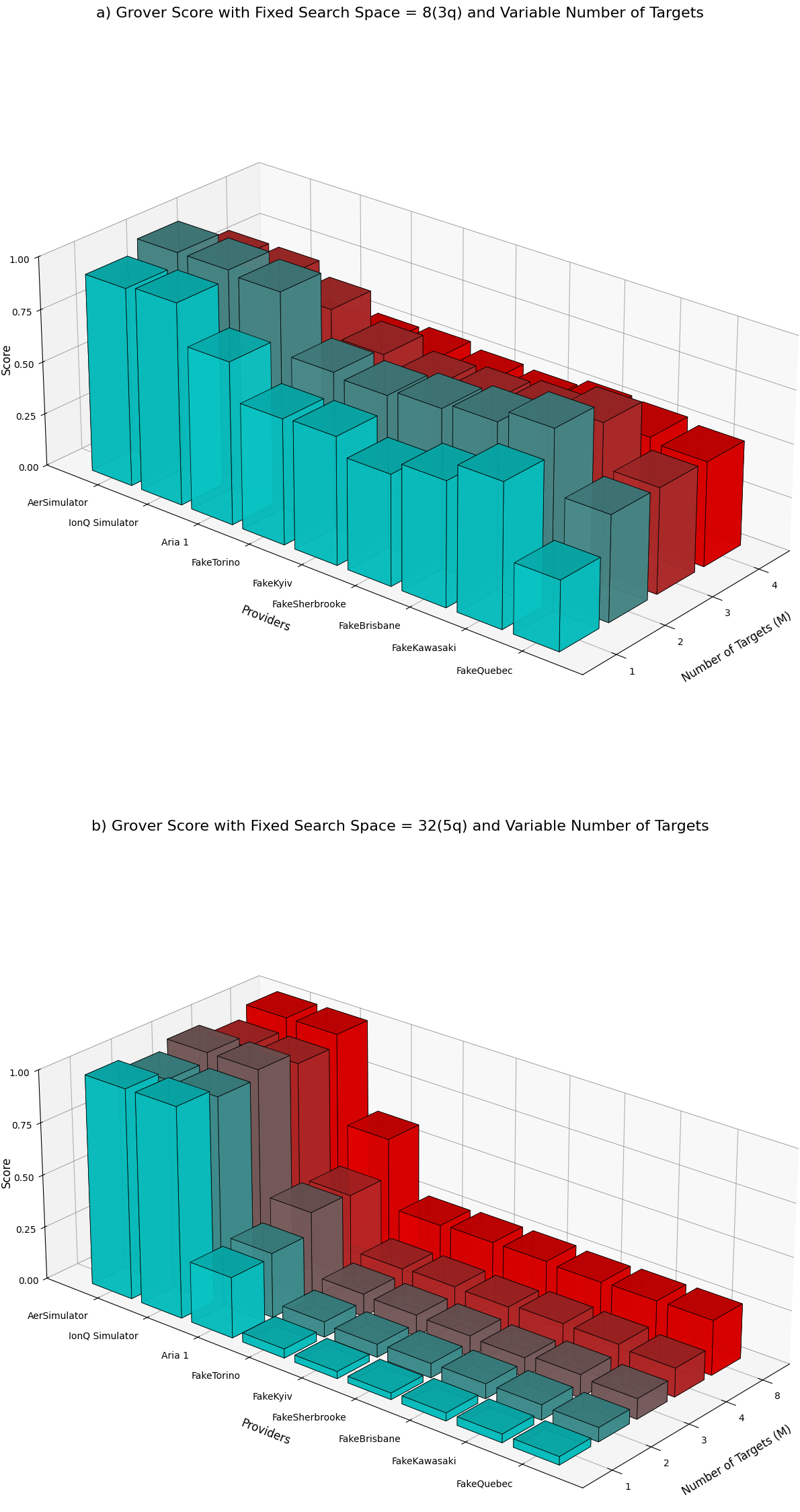}
    \caption{Comparative benchmarking of the \textit{Grover Score} with a variable number of targets. \textbf{a)} For $N=8$, a performance peak is observed at $M=2$ coinciding with the optimal theoretical rotation, and generalized degradation to $\sim 0.5$ at $M=4$ due to Grover's algorithmic limit ($M = N/2$). \textbf{b)} For $N=32$, ideal simulators maintain high fidelity, while \textit{Fake Providers} suffer severe decoherence, relegating their scores to the probability of classical random guessing ($M/N$). \texttt{Aria 1} demonstrates greater resilience to noise, significantly surpassing IBM models in this search space.}
    \label{fig:grover_targets_general}
\end{figure}

In this experiment, we evaluated the response of the \textit{score} by varying the solution density in two fixed search spaces: $N=8$ ($n=3$ qubits) and $N=32$ ($n=5$ qubits). For the first case, the number of target states was incrementally increased as $M \in \{1, 2, 3, 4\}$, while for the second we evaluated $M \in \{1, 2, 3, 4, 8\}$. We contrasted the performance of ideal simulators (\texttt{AerSimulator}, \texttt{IonQ Simulator}) against noise models based on real processors (the IBM \textit{Fake Providers} family and the IonQ \texttt{Aria 1} model). Figure~\ref{fig:grover_targets_general} illustrates the resulting \textit{scores} for both scenarios, highlighting the impact of circuit depth as qubits scale.

The analysis of results reveals a predictable nonlinear behavior. For $N=8$, ideal simulators achieve a perfect score of $1.000$ at $M=2$ (corresponding to the initial optimal amplitude rotation of $\theta = 30^\circ$). IBM architectures suffer noise penalties, varying between $0.495$ and $0.792$, and upon reaching Grover's algorithmic limit at $M=4$ ($M=N/2$), all scores on noisy hardware predictably collapse toward $\approx 0.500$.

On the other hand, for $N=32$ a similar theoretical behavior occurs in equivalent proportions (such as the perfect score at $M=8$), but a much more severe degradation is evident in physical execution: the greater circuit depth for 5 qubits causes IBM models to collapse completely due to decoherence, reducing their results to classical random guessing probabilities ($\approx 0.25$). In this general trade-off of scale and error captured by the metric, IonQ's \texttt{Aria 1} model stands out for having notably superior resilience, being the only one capable of separating from the random limit and preserving quantum advantage at scale.

To isolate the effect of physical topology and evaluate scalability, the second experiment fixed the search to a single solution ($M=1$) and exponentially scaled the size of the search space $N \in \{8, 16, 32, 64, 128\}$, corresponding to circuits from 3 to 7 qubits. The same FoM was executed through the framework in different environments.

\begin{figure}[H]
    \centering
    \includegraphics[width=0.89\linewidth]{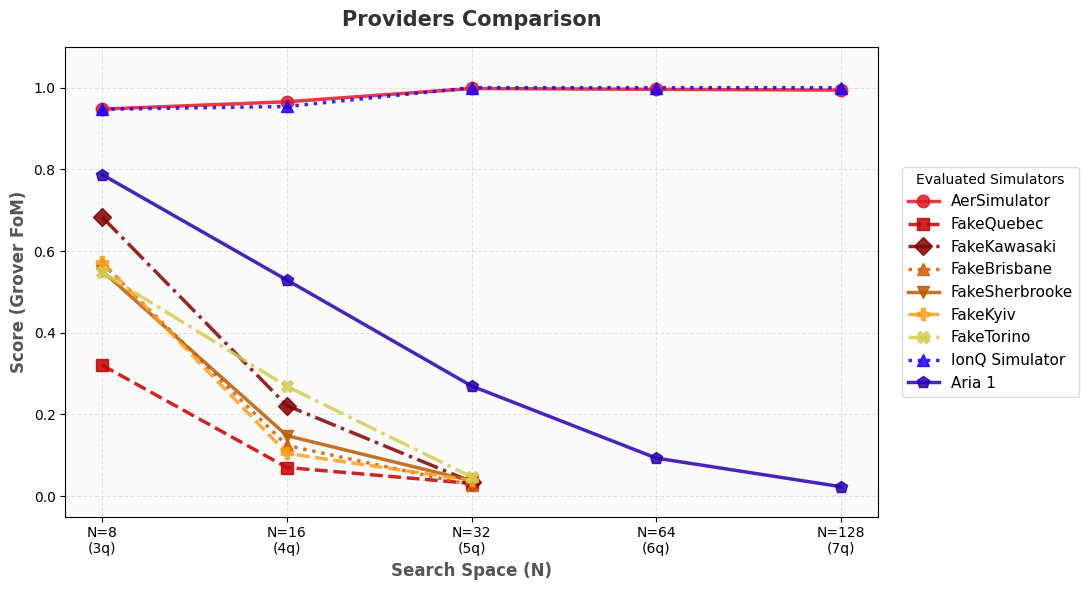}
    \caption{Score comparison across different backends}
    \label{fig:allP}
\end{figure}

First, ideal simulator lines confirm Grover's theoretical limit, converging on practically identical values. As observed in Figure~\ref{fig:allP}, for $N=8$, the ideal simulation yields a score of approximately $0.960$. This is not a simulation artifact, but the manifestation of Grover's deterministic overshoot: forced to apply an integer number of iterations ($R=2$), the quantum state rotates slightly past the orthogonal axis of solutions. Likewise, at $N=32$, $R=4$ iterations fall almost millimetrically on the vertical axis, reaching an asymptotic value of $0.999$. Any value below this baseline mathematically isolates the physical or transpilation impact of hardware.

Finally, a drop in scores is observed in IBM processors as $N$ scales, empirically exposing the cost of topology for highly entangled algorithms. Upon transitioning to $N=32$ (5 qubits), the score collapses to marginal values ($\approx 0.040$). This degradation is explained not only by static gate error rates but also by the decomposition cost originating from the massive insertion of 2-qubit gates necessary to connect distant qubits during the computation of \texttt{mcx} operations. These results demonstrate that Qonscious allows developers to quantify the expected performance of a processor under noise models derived from physical hardware against an algorithmic workload, translating complex topological differences into a unified and comparable score.

\section{Conclusion}
The implementation of GroverFigureOfMerit via polymorphic adapters proves to be an effective solution for addressing heterogeneity in the NISQ era. By abstracting API peculiarities and native gate sets, FoMs enable direct and standardized comparisons between providers with distinct physical architectures. The results confirm that the proposed metric is sensitive to noise and topology differences between backends, demonstrating its ability to discriminate under realistic noise conditions derived from physical processors. This validates the feasibility of evaluating heterogeneous processors under a common metric and facilitates informed backend selection. Validation on real QPUs constitutes the natural confirmation step for these findings. Beyond mere characterization, FoMs in Qonscious optimize resources by conditioning execution at runtime, which can significantly reduce costs in cloud quantum computing environments. The source code and reproducible experiments are available at \url{https://github.com/lifia-unlp/qonscious}, inviting the community to adopt and validate this approach on real hardware.

It is important to note that the choice of Grover's algorithm strictly adheres to its utility as a characterization tool (\textit{stress test}) and not in pursuit of algorithmic advantage. While the debate over the practical utility and advantage of Grover in the short term on noisy hardware remains open, its dense structure of multi-controlled oracles and diffusion operators makes it a useful instrument for testing routing capabilities and hardware coherence, regardless of its current viability for solving real-world search problems. However, experimentation also exposed a bottleneck in the quantum ecosystem: the cost of transpilation. Currently, delegating the decomposition of complex gates (such as \texttt{mcx}) to each provider's default compilers generates severe topological depth overhead, especially in architectures with restricted connectivity. While it would be possible to mitigate this by manually encoding the algorithm using the optimal native gates for each specific provider, this approach breaks the agnostic paradigm, nullifies code scalability, and is highly costly in terms of development time. This underscores the need for smarter abstractions that mediate between the logical circuit and physical constraints.

\section{Future Work}
Building on these findings, the following directions for future work are outlined. First, investigating the development or integration of an intermediate representation (IR) or an agnostic transpilation module would enable optimized translation to native gates, minimizing topological cost across the board without requiring developers to write provider-specific code. On the other hand, validation on real physical processors would allow contrasting results obtained under derived noise models with the actual dynamic behavior of hardware, including phenomena such as \textit{crosstalk}, thermal fluctuations, and variability between executions that static models do not fully capture. Finally, building on the positive results obtained, expanding the FoM catalog toward algorithms with entanglement patterns distinct from Grover's is imperative. Implementing FoMs based on variational hybrid algorithms (VQE, QAOA) or in subroutines with demonstrated advantage such as the Quantum Fourier Transform (QFT) would allow evaluating whether the agnostic approach holds up against diverse circuit structures, while also consolidating a more accurate metric of each architecture's true quantum acceleration potential.

\printbibliography

@article{lammers2025,
  author  = {Lammers, Marcos Guillermo and Holik, Federico Hernán and Fernández, Alejandro},
  title   = {Quantum Resource Management in the NISQ Era: Challenges, Vision, and a Runtime Framework},
  year    = {2025},
  journal = {arXiv preprint arXiv:2508.19276},
  doi     = {10.48550/arXiv.2508.19276},
  url     = {https://doi.org/10.48550/arXiv.2508.19276}
}

@article{manor2025,
  author  = {Manor, Shay and Kumar, Millan and Behera, Priyank and Khalid, Azain and Zeng, Oliver},
  title   = {GRADE: Grover-based Benchmarking Toolkit for Assessing Quantum Hardware},
  year    = {2025},
  journal = {arXiv preprint arXiv:2504.19387},
  doi     = {10.48550/arXiv.2504.19387},
  url     = {https://doi.org/10.48550/arXiv.2504.19387}
}

@article{proctor2024,
  author  = {Proctor, Timothy and Young, Kevin and Baczewski, Andrew D. and Blume-Kohout, Robin},
  title   = {Benchmarking quantum computers},
  journal = {Nature Reviews Physics},
  volume  = {7},
  pages   = {105--118},
  year    = {2025},
  doi     = {10.1038/s42254-024-00796-z},
  url     = {https://doi.org/10.1038/s42254-024-00796-z}
}

@inproceedings{grover1996,
  author    = {Grover, Lov K.},
  title     = {A fast quantum mechanical algorithm for database search},
  booktitle = {Proceedings of the 28th Annual ACM Symposium on Theory of Computing (STOC '96)},
  year      = {1996},
  pages     = {212--219},
  publisher = {Association for Computing Machinery},
  address   = {New York, NY, USA},
  doi       = {10.1145/237814.237866},
  eprint    = {quant-ph/9605043},
  archivePrefix = {arXiv}
}

@ARTICLE{lubinski2023,
  author={Lubinski, Thomas and Johri, Sonika and Varosy, Paul and Coleman, Jeremiah and Zhao, Luning and Necaise, Jason and Baldwin, Charles H. and Mayer, Karl and Proctor, Timothy},
  journal={IEEE Transactions on Quantum Engineering}, 
  title={Application-Oriented Performance Benchmarks for Quantum Computing}, 
  year={2023},
  volume={4},
  number={},
  pages={1-32},
  doi={10.1109/TQE.2023.3253761}}

@misc{gonski2026,
      title={Machine Learning on Heterogeneous, Edge, and Quantum Hardware for Particle Physics (ML-HEQUPP)}, 
      author={Julia Gonski and Jenni Ott and Shiva Abbaszadeh and Sagar Addepalli and Matteo Cremonesi and Jennet Dickinson and Giuseppe Di Guglielmo and Erdem Yigit Ertorer and Lindsey Gray and Ryan Herbst and Christian Herwig and Tae Min Hong and Benedikt Maier and Maryam Bayat Makou and David Miller and Mark S. Neubauer and Cristián Peña and Dylan Rankin and Seon-Hee and Seo and Giordon Stark and Alexander Tapper and Audrey Corbeil Therrien and Ioannis Xiotidis and Keisuke Yoshihara and G Abarajithan and Sagar Addepalli and Nural Akchurin and Carlos Argüelles and Saptaparna Bhattacharya and Lorenzo Borella and Christian Boutan and Tom Braine and James Brau and Martin Breidenbach and Antonio Chahine and Talal Ahmed Chowdhury and Yuan-Tang Chou and Seokju Chung and Alberto Coppi and Mariarosaria D'Alfonso and Abhilasha Dave and Chance Desmet and Angela Di Fulvio and Karri DiPetrillo and Javier Duarte and Auralee Edelen and Jan Eysermans and Yongbin Feng and Emmett Forrestel and Dolores Garcia and Loredana Gastaldo and Julián García Pardiñas and Lino Gerlach and Loukas Gouskos and Katya Govorkova and Carl Grace and Christopher Grant and Philip Harris and Ciaran Hasnip and Timon Heim and Abraham Holtermann and Tae Min Hong and Gian Michele Innocenti and Koji Ishidoshiro and Miaochen Jin and Jyothisraj Johnson and Stephen Jones and Andreas Jung and Georgia Karagiorgi and Ryan Kastner and Nicholas Kamp and Doojin Kim and Kyoungchul Kong and Katie Kudela and Jelena Lalic and Bo-Cheng Lai and Yun-Tsung Lai and Tommy Lam and Jeffrey Lazar and Aobo Li and Zepeng Li and Haoyun Liu and Vladimir Lončar and Luca Macchiarulo and Christopher Madrid and Benedikt Maier and Zhenghua Ma and Prashansa Mukim and Mark S. Neubauer and Victoria Nguyen and Sungbin Oh and Isobel Ojalvo and Hideyoshi Ozaki and Simone Pagan Griso and Myeonghun Park and Christoph Paus and Santosh Parajuli and Benjamin Parpillon and Sara Pozzi and Ema Puljak and Benjamin Ramhorst and Amy Roberts and Larry Ruckman and Kate Scholberg and Sebastian Schmitt and Noah Singer and Eluned Anne Smith and Alexandre Sousa and Michael Spannowsky and Sioni Summers and Yanwen Sun and Daniel Tapia Takaki and Antonino Tumeo and Caterina Vernieri and Belina von Krosigk and Yash Vora and Linyan Wan and Michael H. L. S. Wang and Amanda Weinstein and Andy White and Simon Williams and Felix Yu},
      year={2026},
      eprint={2602.22248},
      archivePrefix={arXiv},
      primaryClass={physics.ins-det},
      url={https://arxiv.org/abs/2602.22248}, 
}

@article{jiang2025,
title={Advancements in superconducting quantum computing},
author={Jiang, Yao-Yao and Deng, Chunqing and Fan, Heng and Li, Bing-Yang and Sun, Luyan and Tan, Xin-Sheng and Wang, Weiting and Xue, Guang-Ming and Yan, Fei and Yu, Hai-Feng and Zhang, Ying-Shan and Zhang, Yu-Ran and Zou, Chang-Ling},
journal={National Science Review},
volume={12},
number={8},
pages={nwaf246},
year={2025},
doi={10.1093/nsr/nwaf246}
}

@article{Nation2025,
  author = {Nation, Paul D. and Saki, Abdullah Ash and Brandhofer, Sebastian and Bello, Luciano and Garion, Shelly and Treinish, Matthew and Javadi-Abhari, Ali},
  title = {Benchmarking the performance of quantum computing software for quantum circuit creation, manipulation and compilation},
  journal = {Nature Computational Science},
  volume = {5},
  number = {5},
  pages = {427--435},
  year = {2025},
  doi = {10.1038/s43588-025-00792-y},
  url = {https://doi.org/10.1038/s43588-025-00792-y},
  publisher = {Springer Nature}
}

@online{api_ibm_2026,
  author  = {{IBM Quantum}},
  title   = {Qiskit API Reference: qiskit.providers.BackendV2},
  year    = {2026},
  url     = {https://quantum.cloud.ibm.com/docs/api/qiskit/qiskit.providers.BackendV2},
  urldate = {2026-03-13}
}

@online{api_rigetti_2026,
  author  = {{Rigetti Computing}},
  title   = {Quil-T Getting Started: get\_calibration\_program()},
  year    = {2026},
  url     = {https://pyquil-docs.rigetti.com/en/v4/quilt_getting_started.html},
  urldate = {2026-03-13}
}

@online{api_quantinuum_2026,
  author  = {{Quantinuum}},
  title   = {Quantinuum Nexus User Guide: Backend Snapshots},
  year    = {2026},
  url     = {https://docs.quantinuum.com/nexus/user_guide/concepts/backend_snapshots.html},
  urldate = {2026-03-13}
}

@online{api_ionq_2026,
  author  = {{IonQ}},
  title   = {Using Native Gates with Qiskit},
  year    = {2026},
  url     = {https://www.ionq.com/resources/using-native-gates-with-qiskit},
  urldate = {2026-03-13}
}

@online{api_quera_2026,
  author  = {{QuEra Computing}},
  title   = {Bloqade Reference: Hardware Capabilities},
  year    = {2026},
  url     = {https://bloqade.quera.com/latest/analog/reference/hardware-capabilities/},
  urldate = {2026-03-13}
}

@online{api_xanadu_2026,
  author  = {{Xanadu}},
  title   = {Xanadu Cloud Client (XCC) Documentation},
  year    = {2026},
  url     = {https://xanadu-cloud-client.readthedocs.io/},
  urldate = {2026-03-13}
}
\end{document}